# Measuring thrust and predicting trajectory in model rocketry

Michael Courtney and Amy Courtney
Ballistics Testing Group, P.O. Box 24, West Point, NY 10996
Michael_Courtney@alum.mit.edu

**Abstract:** Methods are presented for measuring thrust using common force sensors and data acquisition to construct a dynamic force plate. A spreadsheet can be used to compute trajectory by integrating the equations of motion numerically. These techniques can be used in college physics courses, and have also been used with high school students concurrently enrolled in algebra 2.

## I. Introduction

Model rocketry generates excitement and enthusiasm. However, many teachers aspire to impart quantitative understanding beyond the initial excitement of rocketry and qualitative understanding of Newton's third law. This paper presents a relatively simple experimental method for measuring a rocket motor's thrust curve and a theoretical approach for predicting the resulting trajectory that has been successfully implemented by high-school students concurrently enrolled in algebra 2. There are a number of excellent resources for hobbyists and teachers entering the field of rocketry.[1,2,3,4,5]

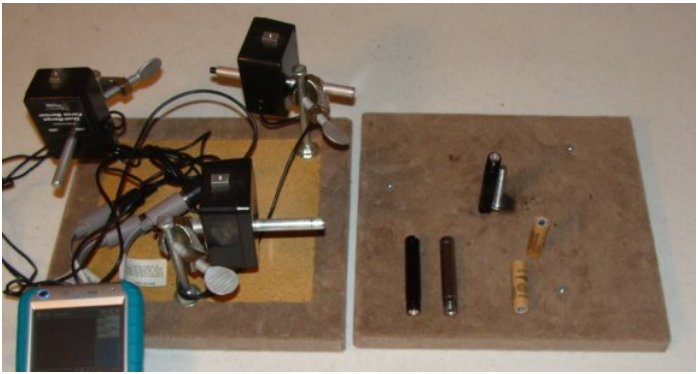

*Figure 1: Force plate for measuring rocket engine thrust curves. The rocket motor is attached to the bolt extending out of the plate on the right which is set on top of the three force sensors on the left. (The photo shows one aluminum cased rocket motor attached to the bolt and two other aluminum cased motors in on the lower left corner of the plate, as well as two brown paper cased Estes A10-PT rocket motors.)*

## II. Experimental method

A number of educational and engineering instrument companies offer solutions for dynamic force measurements that can be successfully adapted for model rocketry. Here we use equipment that is readily available, can be configured with relative ease, and can be used with a software interface that is likely to be integrated with other science experiments so that the test system does not represent an entirely new learning curve. We also wanted a design that is easily calibrated and has sufficient dynamic range to yield accurate results with the smallest model rocket engines and handle thrusts up to 100 N.

The force plate employs three Vernier "dual-range force sensors" connected to a Vernier LabQuest. The three force sensors are attached to a bottom plate as shown in Figure 1, and a force plate rests on top of them and has a bolt to which the rocket engine is attached for static thrust testing. (The plate is held in place by gravity, there is no adhesive or connectors.) The total thrust is the sum of the three individual force readings. (The force plate is zeroed after the motor is attached.) Each force sensor has a selectable range of either *10N* or *50N*, so that if the plate was perfectly balanced, the full scale would be either *30N* or *150N* minus the static load, depending on the sensor setting. However, since the plate is not perfectly balanced, the three force readings are not equal, and the full scale ranges are closer to *20N* and *100N* for the *10N* and *50N* sensor settings, respectively. This system design is capable of measuring thrust curves for the full range (1/4A to E) of commonly available hobby rocket engines, as well as many experimental rocket motor designs.

The LabQuest can be configured for a variety of sample rates. The Logger Pro software both handles calibration and allows the three individual forces to be added and graphed as a total force, as well as reporting the total impulse (integrated area under the force curve). The time delay can also be subtracted from the time base to set the rocket ignition to *t = 0 s*.

Figure 2 shows the measured thrust curve for an Estes A10-PT model rocket engine. The shape of the thrust curve compares favorably with the published curve from the manufacturer.[6] However, our experimental curve has a peak of *11.5 N* and a total impulse of *1.991 Ns*, compared with the manufacturer's claims of a peak thrust of *13.0 N* and a total impulse of *2.5 Ns*. The total impulse (area under total thrust curve) is *20%* lower than the manufacturer's claim. The bulk of this discrepancy is likely due to the engine containing *19% less* (*3.08 g*) than the specified quantity (*3.78 g*) of propellant, since our measured specific impulse (impulse divided by propellant weight) is *64.64 s*, which is only *4%* lower than the manufacturer's specification of *67.49 s* for specific impulse. (Measurements showing the A10-PT being below specification have been consistently repeated on different days using independent calibrations.)

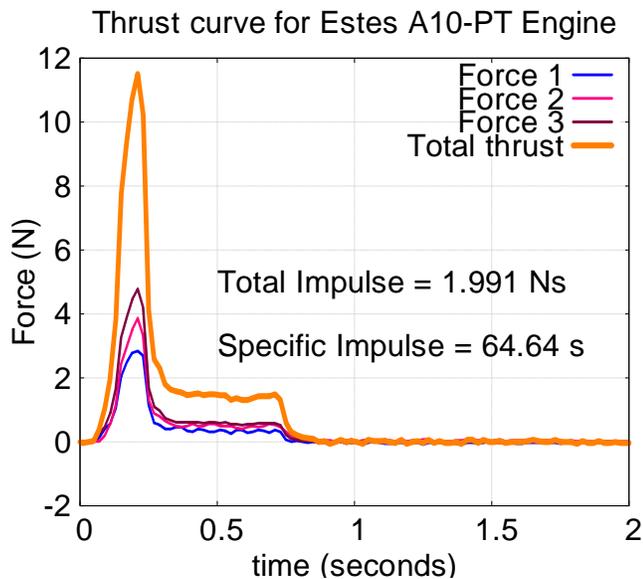

Figure 2: *Thrust curve for Estes A10-PT Engine. This is a relatively small rocket motor (8.08 g total mass, 3.08 g of fuel) capable of reaching heights of over 200m (according to the manufacturer) in some rocket assemblies.*

Figure 3 shows the measured thrust curve for an experimental motor with a sugar-based propellant containing a *65/35/2* blend of $KNO_3$, $C_6H_{12}O_6$, and $Fe_2O_3$. The theoretical specific impulse of this blend can be much higher (above *150 s*)[7] but this motor's operating pressure is relatively low because of the propellant's slow burn rate and the motor's relatively large nozzle size. Filled to capacity (*10 g*) with a sucrose-based propellant, we believe this re-loadable Maglite-based motor can have a specific impulse close to *100 Ns*.

### III. Predicting Trajectory

Since the mass of the rocket is changing during the burn phase of flight, the acceleration is given by

$$a = \frac{F_{thrust} - F_{weight} - F_{drag} - V\frac{dm}{dt}}{m},$$

where $V$ is the instantaneous velocity, and $m$ is the mass. Setting up the spreadsheet requires estimating the mass as a function of time. A linear interpolation between the initial and final masses can be used. Since the changing mass term and weight are much smaller than the thrust, and also smaller than the drag, there is only a small error from the linear interpolation.

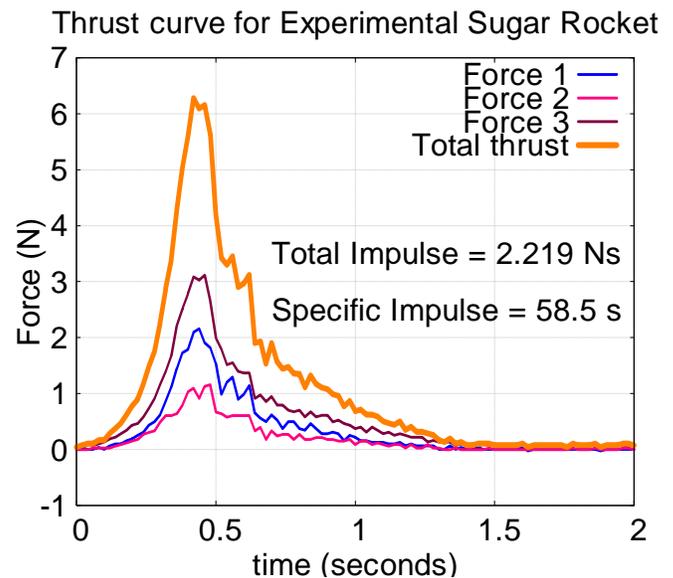

Figure 3: *Thrust curve for experimental sugar rocket using a 65/35/2 blend of potassium nitrate, dextrose, and red iron oxide in a re-loadable motor casing made from a Maglite Solitaire aluminum flashlight case. This trial used 3.87 g of propellant, a nozzle with a 0.125" throat diameter, and a propellant grain with a 0.125" diameter core.*

The drag force is $F_{drag} = -\frac{1}{2}C_d \cdot A \cdot \rho \cdot V^2$, where $C_d$ is the drag coefficient (unitless), $A$ is the frontal area of the rocket ($m^2$), $\rho$ is the density of air ($kg/m^3$), and $V$ is the rocket velocity (*m/s*). Care is required with the signs (adding or subtracting) in the spreadsheet. The weight is always subtracted, but the drag is always in the opposite direction of motion. The changing mass term enters the equation with a negative sign, but effectively increases the acceleration (a positive pseudo-force) because the change in mass is negative.

Each row of the spreadsheet corresponds an instant of time. Each column corresponds to a physical quantity: time, thrust, mass, $-V(dm/dt)$, weight, drag, net force, acceleration, velocity, height, and impulse. During the thrust phase, the time step is *0.02 s*. Values are computed working from left to right in each successive row. The thrust was obtained with the force plate described above. A linear interpolation is used to estimate the changing mass. $-V(dm/dt)$ is computed using the velocity from the previous row and the change in mass from the previous row to the current row. The weight is mg. The drag force is computed using the velocity from the previous row. The acceleration is then computed using values from the current row. Once each acceleration is computed, each velocity and position are computed from the kinematic equations, $v = v_i + a\,dt$ and $y = y_i + v\,dt$, where $dt$ is the time interval, *0.02 s*. $v_i$ and $y_i$ are the velocity and height from the previous row. The impulse is computed for each time interval as $F_{thrust}\,dt$, and the column is summed after the fuel is all consumed. Shortly after fuel burnout, the step size can be increased to *0.1 s* because the forces are changing slowly for the duration of the flight.

The forces, velocity, and height curves are shown in Figures 4, 5, and 6 for the Estes A10-PT motor. The trajectory has three distinct phases: the thrust phase from *0.0 s* to *0.8 s*. This phase ends when the thrust falls to zero and the net force suddenly becomes negative. No longer dominated by the thrust, the net force is then dominated by air drag and rocket weight, which are both negative and rapidly robbing the rocket of velocity ($a = -25.88\ m/s^2$ just after burnout). After burnout, the rocket continues ascending during the coast phase until approximately $t = 5.0\ s$ when the rocket reaches its peak because drag and weight have reduced the velocity to zero. After reaching the peak, the rocket falls in the descent phase. Care is required when entering the drag formula so that it will be upward (change directions) after the peak. Rather than squaring the velocity, the $V^2$ term is expressed as $V|V|$ to ensure that the drag force properly changes sign when the rocket begins to descend. The rocket returns to the ground at approximately $t = 10.5s$.

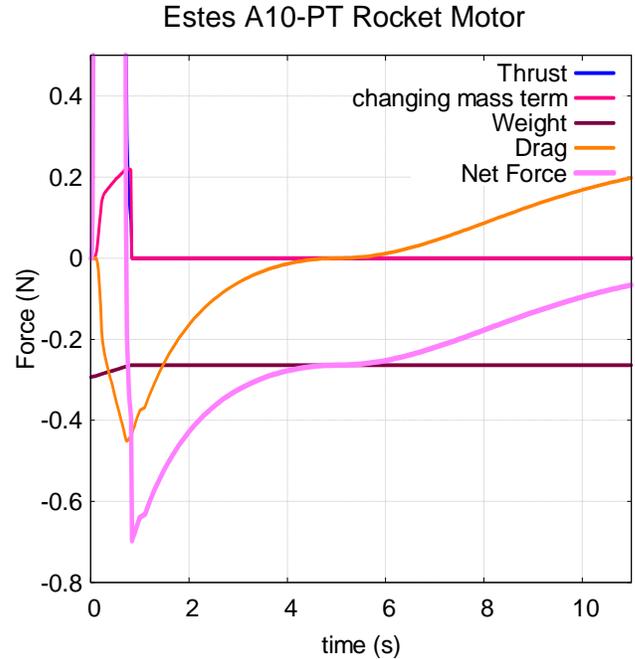

*Figure 4: Forces on a rocket using an Estes A10-PT rocket motor. These calculations use a total initial mass of 30 g, a drag coefficient of 0.4, an air density of 1.29 kg/m³, and a frontal diameter of 2.54 cm.*

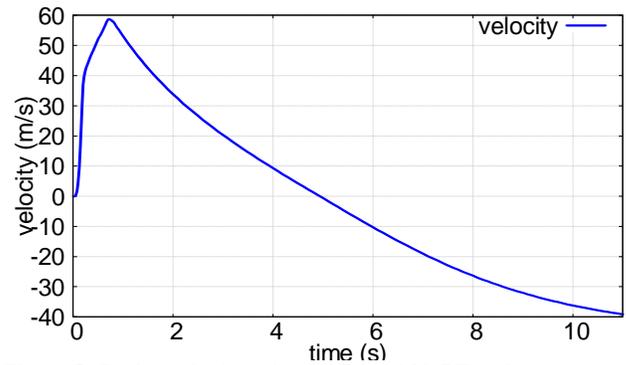

*Figure 5: Rocket velocity using an Estes A10-PT rocket motor.*

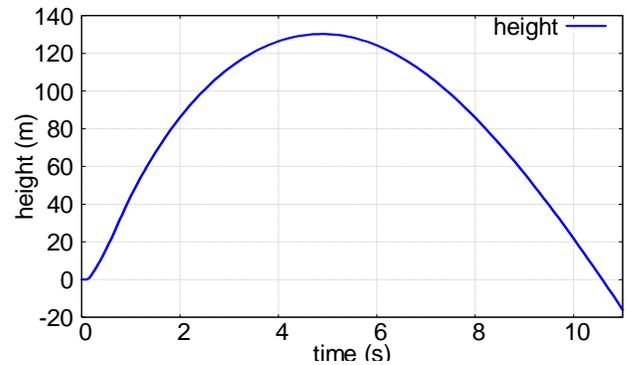

*Figure 6: Rocket height using an Estes A10-PT rocket motor.*

## IV. Discussion

A number of simplifying assumptions have been made to make the problem solvable and tractable with algebra 2 level skills. First, we assume a straight upward launch with no deviation from vertical. Second, we model the changing mass with a linear interpolation. Third, typical values are used for the coefficient of drag and the density of air. Knowing these values with confidence requires additional considerations.

However, even with these simplifying assumptions, the rocket trajectory computation here goes significantly beyond the typical introductory projectile problem where the force is constant, mass is constant, and drag is neglected. The approach presented here allows for modeling these effects with reasonable accuracy and is a significant step above the qualitative description that a rocket's motion is merely the equal and opposite reaction to the expanding gases expelled from the nozzle.

The force plate demonstrated here can also be useful for evaluating experimental rocket motor designs including propellant grain shape, nozzle design, propellant chemistry, and grain production techniques. Each propellant has a theoretical specific impulse which can be computed from the theory of rocketry[8] or using widely available software such as PROPEP and GUIPEP.[9] Comparing the theoretical specific impulse with the experimentally determined value indicates the combustion pressure and the efficiency of the rocket motor design (grain, nozzle, etc.)

## V. Appendix

The first ten rows of the spreadsheet predicting the trajectory of a *30 g* rocket using an Estes A10-PT engine are shown in the table below. Using a step size of *0.02 s* from *t=0 s* to *t=1.0 s* and a step size of *0.1 s* from *t=1.0s* to *t = 11.0s* requires *151* lines total and yields the predictions shown in Figures 4, 5, and 6.

| Time (s) | Thrust (N) | Mass (g) | -V(dm/dt) (N) | Weight (N) | Drag (N) | Fnet (N) | a (m/s$^2$) | V (m/s) | height (m) | Impulse (Ns) |
|---|---|---|---|---|---|---|---|---|---|---|
| 0.0000 | 0.0000 | 30.0000 | 0.0000 | 0.2940 | 0.0000 | 0.0000 | 0.0000 | 0.0000 | 0.0000 | 0.0000 |
| 0.0200 | 0.0305 | 29.9249 | 0.0000 | 0.2933 | 0.0000 | 0.0000 | 0.0000 | 0.0000 | 0.0000 | 0.0006 |
| 0.0400 | 0.3964 | 29.8498 | 0.0000 | 0.2925 | 0.0000 | 0.1039 | 3.4805 | 0.0696 | 0.0014 | 0.0079 |
| 0.0600 | 1.0633 | 29.7746 | 0.0003 | 0.2918 | 0.0000 | 0.7717 | 25.9194 | 0.5880 | 0.0132 | 0.0213 |
| 0.0800 | 2.0076 | 29.6995 | 0.0022 | 0.2911 | 0.0000 | 1.7187 | 57.8706 | 1.7454 | 0.0481 | 0.0402 |
| 0.1000 | 3.8322 | 29.6244 | 0.0066 | 0.2903 | -0.0004 | 3.5480 | 119.7676 | 4.1408 | 0.1309 | 0.0766 |
| 0.1200 | 7.7592 | 29.5493 | 0.0156 | 0.2896 | -0.0022 | 7.4830 | 253.2370 | 9.2055 | 0.3150 | 0.1552 |
| 0.1400 | 9.3685 | 29.4742 | 0.0346 | 0.2888 | -0.0111 | 9.1032 | 308.8531 | 15.3826 | 0.6226 | 0.1874 |
| 0.1600 | 10.7396 | 29.3990 | 0.0578 | 0.2881 | -0.0309 | 10.4783 | 356.4158 | 22.5109 | 1.0729 | 0.2148 |
| 0.1800 | 11.5009 | 29.3239 | 0.0846 | 0.2874 | -0.0662 | 11.2318 | 383.0250 | 30.1714 | 1.6763 | 0.2300 |
| 0.2000 | 10.2216 | 29.2488 | 0.1133 | 0.2866 | -0.1190 | 9.9293 | 339.4776 | 36.9609 | 2.4155 | 0.2044 |